\documentclass[ALICE,manyauthors,nocleardouble]{cernphprep}
\usepackage[utf8]{inputenc}
\usepackage[comma,square,numbers,sort&compress]{natbib}
\usepackage[hyperfootnotes=false]{hyperref}
\usepackage{mathtools}
\usepackage[capitalise]{cleveref}
\usepackage{amsmath}
\usepackage{bm}
\usepackage{color}
\usepackage{graphicx}
\usepackage{siunitx}
\usepackage{float}
\usepackage{microtype,relsize}
\usepackage{placeins}
\xspaceaddexceptions{]\}}
\usepackage{lineno}
\DeclarePairedDelimiterX\avg[1]{\langle}{\rangle}{#1}

\newcommand{\vt}[1]{\ensuremath{v_{2}\left\{#1\right\}}\xspace}
\newcommand{\PsiX}[1]{\ensuremath{\Psi_{\mathrm{#1}}}\xspace}

\newcommand{\vn}{\ensuremath{v_{n}}\xspace}
\newcommand{\dNdphi}{\ensuremath{\mathrm{d}N/\mathrm{d}\varphi}}
\newcommand{\PsiPP}{\PsiX{n}}
\newcommand{\PsiSP}{\PsiX{\mathrm{SP}}}
\newcommand{\PsiSPT}{\ensuremath{\PsiX{\mathrm{SP}}^{\mathrm{T}}}\xspace}
\newcommand{\PsiSPP}{\ensuremath{\PsiX{\mathrm{SP}}^{\mathrm{P}}}\xspace}
\newcommand{\PsiRP}{\PsiX{\mathrm{RP}}}
\newcommand{\vtsp}{\vt{\PsiSP}}

\newcommand{\vtetagap}{\vt{2, \left|\Delta\eta\right| > 1}}
\newcommand{\vtetagapnoone}{\vt{2, \left|\Delta\eta\right|}}
\newcommand{\vtf}{\vt{4}}
\newcommand{\vtt}{\vt{2}}
\newcommand{\vtspvtf}{\ensuremath{\left. \vtsp \right/ \vtf}\xspace}

\newcommand{\eprp}{\ensuremath{\varepsilon_{\mathrm{RP}}}\xspace}
\newcommand{\eptf}{\ensuremath{\varepsilon_{2}\{4\}}\xspace}
\newcommand{\eptt}{\ensuremath{\varepsilon_{2}\{2\}}\xspace}

\newcommand{\Epbpb}{{\ensuremath{\sqrt{s_{\mathrm{NN}}} = \SI{2.76}{TeV}}}\xspace}
\newcommand{\Exexe}{{\ensuremath{\sqrt{s_{\mathrm{NN}}} = \SI{5.44}{TeV}}}\xspace}

\newcommand{\TRENTo}{T\textsubscript{R}ENTo\xspace}

\DeclareSIUnit\clight{\textit{c}}
\DeclareSIUnit\eVperc{\eV\per\clight}
\DeclareSIUnit\GeVc{\giga\eV\per\clight}
\DeclareSIUnit\GeV{\giga\eV}
\DeclareSIUnit\fmc{\femto\m\per\clight}
\DeclareSIUnit\fm{\femto\m}
\DeclareSIUnit\TeV{\tera\eV}
\DeclareSIUnit\MeVc{\mega\eV\per\clight}
\DeclareSIUnit\barn{b}
\DeclareSIUnit\kilo{k}

\newcommand{\sqrts}{\ensuremath{\sqrt{s_{\mathrm{NN}}}}}

\newcommand{\dNchdEta}{\ensuremath{\mathrm{d}N_{\mathrm{ch}}/\mathrm{d}\eta}}

\newcommand{\pt}{\ensuremath{p_{\mathrm{T}}}\xspace}

\newcommand{\PbPb}{Pb--Pb\xspace}
\newcommand{\XeXe}{Xe--Xe\xspace}
\usepackage{transparent}
\graphicspath{Figures}
\usepackage[T1]{fontenc}
\usepackage{orcidlink}

\begin{document}%
\begin{titlepage}
\PHyear{2022}       
\PHnumber{061}      
\PHdate{24 March}   
\title{Elliptic flow of charged particles at midrapidity relative to the spectator plane in \PbPb and \XeXe collisions}
\ShortTitle{Elliptic Flow Fluctuations at Midrapidity using Spectator Neutrons}
 \Collaboration{ALICE Collaboration\thanks{See Appendix~\ref{app:collab} for the list of collaboration members}}
\ShortAuthor{ALICE Collaboration}
\begin{abstract}
Measurements of the elliptic flow coefficient relative to the collision plane defined by the spectator neutrons \vtsp in collisions of Pb ions at center-of-mass energy per nucleon--nucleon pair \Epbpb and Xe ions at \Exexe are reported.
The results are presented for charged particles produced at midrapidity as a function of centrality and transverse momentum for the 5--70\% and \SIrange{0.2}{6}{\GeVc} ranges, respectively.
The ratio between \vtsp and the elliptic flow coefficient relative to the participant plane \vtf, estimated using four-particle correlations, deviates by up to 20\% from unity depending on centrality.
This observation differs strongly from the magnitude of the corresponding eccentricity ratios predicted by the \TRENTo and the elliptic power models of initial state fluctuations that are tuned to describe the participant plane anisotropies.
The differences can be interpreted as a decorrelation of the neutron spectator plane and the reaction plane because of fragmentation of the remnants from the colliding nuclei, which points to an incompleteness of current models describing the initial state fluctuations.
A significant transverse momentum dependence of the ratio \vtspvtf is observed in all but the most central collisions, which may help to understand whether momentum anisotropies at low and intermediate transverse momentum have a common origin in initial state fluctuations.
The ratios of \vtsp and \vtf to the corresponding initial state eccentricities for \XeXe and \PbPb collisions at similar initial entropy density show a difference of $(7.0\pm0.9)\%$ with an additional variation of $+1.8\%$ when including RHIC data in the \TRENTo parameter extraction.
These observations provide new experimental constraints for viscous effects in the hydrodynamic modeling of the expanding quark--gluon plasma produced in heavy-ion collisions at the LHC.
\end{abstract}
\end{titlepage}

Determining the properties of deconfined quark--gluon matter, called quark--gluon plasma (QGP), is the goal of the heavy-ion program at the Large Hadron Collider (LHC).
Previous studies of data collected at the Brookhaven Relativistic Heavy-Ion Collider (RHIC) and the LHC have shown that the QGP behaves like a liquid with very small specific shear and bulk viscosities~\cite{Ryu:2015vwa, Schenke:2010rr}.
In heavy-ion collisions, these properties are encoded in the collective expansion of the strongly interacting QGP.
Measurements of this collective behavior, in particular the anisotropic flow driven by the spatial anisotropy of the shape of the overlap region of the colliding nuclei, can be used to infer the QGP transport properties.
The anisotropic flow is quantified by the coefficients $v_n$ of a Fourier decomposition of the momentum anisotropy of emitted particles relative to the collision symmetry planes with angles $\Psi_n$ of the harmonic $n$.
The dominant coefficient in off-center collision of heavy ions is the elliptic flow $v_2$.
The shape of the initial energy density distribution in the overlap region of the nuclei (participant zone) fluctuates from collision to collision due to the motion of nucleons and the quantum mechanical nature of the nucleus--nucleus interaction, which is typically modeled by spatial position fluctuations of the interacting nucleons.
This was initially discussed in the context of the elliptic flow fluctuations~\cite{Miller:2003kd,PHOBOS:2007rsf, Sorensen:2008zk} and later confirmed by observations~\cite{Alver:2010gr,PHENIX:2011yyh,ALICE:2011ab,STAR:2013qio,STAR:2016vqt} of significant non-zero triangular flow $v_3$ and higher flow harmonics.
A detailed picture of fluctuations has already been inferred from the comparison of initial state models to measurements of flow fluctuations relative to the participant symmetry planes~\cite{Yan:2014afa,Moreland:2014oya,ALICE:2021gxt,ALICE:2011ab, ATLAS:2012at,Chatrchyan:2013kba,Aad:2013xma,Aad:2014fla, Aad:2015lwa,ALICE:2016kpq,Adam:2016izf,Sirunyan:2017fts,ALICE:2017lyf,Acharya:2017zfg,Acharya:2018lmh,ALICE:2018lao,Acharya:2019uia,Acharya:2020taj,Acharya:2021hpf}.
Additional information about the pattern of initial state fluctuations can be obtained from measurements using spectator nucleons, which decouple very fast from the participant zone.

\begin{figure}[hb]
\centering
\begin{subfigure}[b]{0.45\textwidth}
\includegraphics[width=1.0\textwidth]{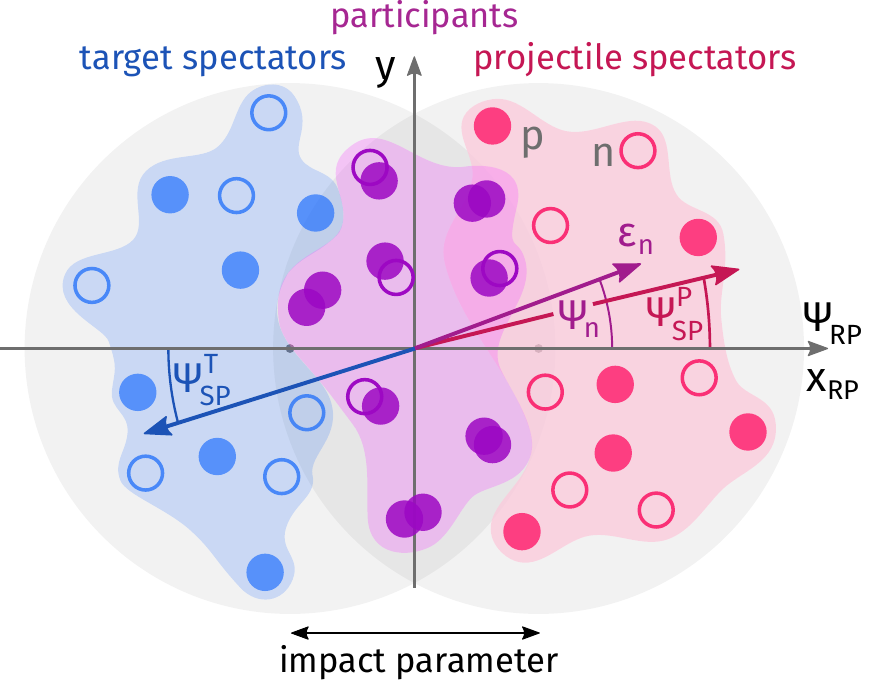}
\caption{transverse view}
\label{fig1a}
\end{subfigure}
\hspace{1em}
\begin{subfigure}[b]{0.45\textwidth}
\includegraphics[width=1.0\textwidth]{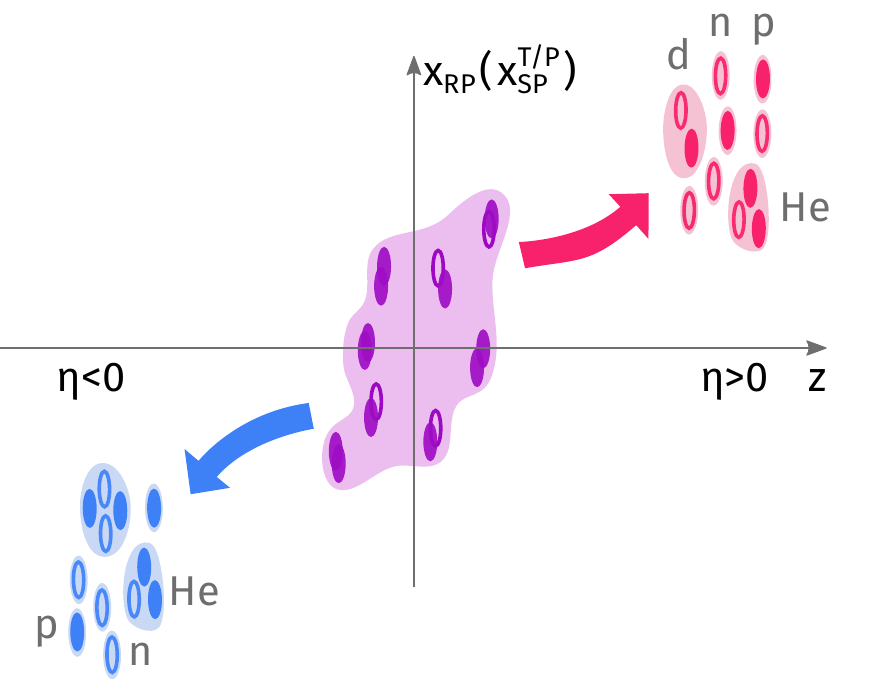}
\caption{longitudinal view}
\label{fig1b}
\end{subfigure}
\caption{
(color online)
A sketch of the geometry of a non-central heavy-ion collision in the \subref{fig1a} transverse $(x_{\mathrm{RP}},y)$ and \subref{fig1b} reaction $(x_{\mathrm{RP}},z)$ planes.
The $x_{\mathrm{RP}}$ axis points along the direction of the impact parameter given by the distance between the centers of the colliding nuclei, while the $z$ axis is oriented along the direction of the colliding nuclei.
The full (open) circles represent the protons (neutrons) from projectile (red) and target (blue) nuclei.
The participant nucleons are shown in purple.
In panel~\subref{fig1a}, the arrows indicate the eccentricity vectors ($\epsilon_n$ is shown as example) and corresponding reaction (\PsiRP), projectile/target spectator ($\Psi_\mathrm{SP}^{\rm P,T}$), and  participant (multiple \PsiPP, $n\ge1$) plane angles.
Panel \subref{fig1b} shows the outward deflection, indicated by curved arrows, of the spectators and fragmentation products of the projectile ($\eta>0$) and target ($\eta<0$) recoil nuclei, such as protons ($p$), neutrons ($n$), and nuclei (deuteron, helium, etc).
}%
\label{fig1}
\end{figure}

\Cref{fig1} sketches the geometry of a non-central heavy-ion collision, divided into the regions of the participant nucleons (i.e. those encountering strong interactions) and the deflected spectator nucleons of the target (T) and projectile (P) nuclei.
The geometrical reaction plane, denoted by its angle \PsiRP in \cref{fig1a}, is spanned by the impact parameter, pointing in the direction of the $\mathrm{x}_{\mathrm{RP}}$ axis, and the movement direction of the colliding nuclei, which is indicated by the z axis.
The fluctuating shape of the energy density distribution of the collision can be characterized by the eccentricities $\varepsilon_n$ for $n>1$
\begin{equation}
    \label{eq:eccentricity}
    \varepsilon_n e^{in\Psi_n} = -\frac{\int r^n e^{in\varphi} \rho(r,\varphi)r\mathrm{d} r \mathrm{d} \varphi}
    {\int r^n \rho(r,\varphi) r \mathrm{d} r \mathrm{d} \varphi}\ ,
\end{equation}
where $\rho(r,\varphi)$ is the initial energy or entropy density near midrapidity and $(r, \varphi)$ are the polar coordinates.
\cref{eq:eccentricity} for $n=2$ gives the standard formula for the ellipticity~\cite{Teaney:2010vd, Yan:2017ivm}.
In the absence of fluctuations, the azimuthal symmetry plane of the collision is given by the reaction plane.
In the presence of fluctuations, the collision symmetry is not described by a single reaction plane, but by multiple participant planes with angles $\Psi_n$ and spectator planes with angles \PsiSPT and \PsiSPP.
The spectator planes are spanned by the beam direction and the deflection of the spectators, which is determined by the sum of the momenta of the spectator nucleons and fragments.
For small fluctuations of the spectator deflection, the target and projectile spectator planes share a common symmetry plane with angle \PsiSP.
The sketch is oriented along the reaction plane $x_{\mathrm{RP}}$ (i.e. $\PsiRP=0$), while in the experiment the orientation of the collision relative to the laboratory frame fluctuates from one collision to another. This causes the reaction plane angle to be different from zero.

\Cref{fig1b} illustrates a longitudinal view of the early time of a heavy-ion collision, in particular the deflection of the spectators.
The fragmentation of the nuclear recoil is a complex process~\cite{Schuttauf:1996ci} involving the release of protons, neutrons, and other nuclei, which causes a decorrelation of the reaction plane and spectator planes.
Additionally, a plane defined only by a subset of the spectator fragments (e.g. only neutrons or protons) can be different from the overall spectator plane orientation illustrated in~\cref{fig1a}.
This difference depends on the collision centrality due to the balance between the energy carried by free nucleons and nuclear fragments.

The deflection of the spectators along the spectator planes at LHC collision energies, as illustrated by the red and blue arrows in \cref{fig1a}, was shown to be outward from the nuclear overlap region, based on the measured directed flow relative to the neutron spectator plane and hydrodynamic model calculations~\cite{Voloshin:2016ppr}.
From measurements of the neutron deflection at the LHC, the transverse momentum exchange between the participants and the spectators has been estimated to be in the order of \SI{20}{\MeVc} per nucleon in midcentral collisions ~\cite{ALICE:2013xri}.
This is about an order of magnitude smaller than the width of the transverse momentum distribution of emitted spectators, which is proportional~\cite{Goldhaber:1974qy,Morrissey:1989wv} to the Fermi momentum of \SI{265}{\MeVc}~\cite{Moniz:1971mt} for lead nuclei.

Due to the geometry of the collision, the direction in which the  spectators are deflected is believed to be strongly correlated to the reaction plane.
In collisions with large impact parameters, the smaller number of sources of particle production (number of participants) results in a stronger decorrelation of the participant symmetry plane and the reaction plane.
Measurements of the directed flow using the deflection of spectator neutrons showed a small but non-zero decorrelation between the direction of the neutron deflection and the reaction plane orientation, in addition to a relative decorrelation of the neutron spectator planes originating from the projectile and target nuclei, respectively~\cite{ALICE:2013xri}.
A strong correlation between the participant and spectator planes is expected in midcentral collisions because of the dominance of the elliptic flow~\cite{ALICE:2011svq}.
Together with measurements relative to the participant planes, a measurement relative to the spectator deflection provides unique information about spatial orientation of the participant zone relative to the geometrical orientation of the colliding nuclei~\cite{ALICE:2013xri}.

Hydrodynamic model calculations with small viscosities indicate an approximately linear response to the initial eccentricities for elliptic and triangular flow~\cite{Qiu:2011iv}.
In ideal hydrodynamics, the elliptic flow $v_2$ is proportional to the initial eccentricity~\cite{Gardim:2011xv}
\begin{equation}
  \label{eq:linear_scaling}
  v_2 \propto \kappa_2 \varepsilon_2\ .
\end{equation}
This has been demonstrated by the comparison of measurements relative to the participant plane with eccentricities from different initial state model calculations~\cite{ALICE:2018lao}.
The scaling coefficient $\kappa_2$ is related to the initial entropy density.
In ideal hydrodynamics, it is estimated from the final state multiplicity $\mathrm{d}N_{\mathrm{ch}}/\mathrm{d}y$ per unit of the overlap area $S$~\cite{Voloshin:1999gs,Song:2008si, Heinz:2013th}
\begin{equation}
  \label{eq:mult_scaling}
  \frac{v_2}{\varepsilon_2} \propto \frac{1}{S} \frac{\mathrm{d}N_{\mathrm{ch}}}{\mathrm{d}y}\ .
\end{equation}
Following Ref.~\cite{ALICE:2018lao}, the area $S$ is defined as $S=4\pi\sigma_\mathrm{x}\sigma_\mathrm{y}$, where $\sigma_{\mathrm{x}}$ is the spatial width of the participant zone along the reaction plane direction ($\mathrm{x_{RP}}$) and $\sigma_{\mathrm{y}}$ is along the transverse direction (y).
This scaling can be tested by comparing different collision systems and their dependence on centrality or center-of-mass energy.
Such a scaling is broken by viscous effects, which depend on the magnitude of the temperature-dependent specific shear and bulk viscosities~\cite{Song:2008si}.
Even for an ideal fluid, the scaling can be broken during the final stages of the evolution before freeze-out~\cite{Song:2008si}.

One of the largest uncertainties in the description of heavy-ion collisions is the pattern of fluctuations in the initial state that results in large uncertainties of the extracted QGP transport properties~\cite{Bernhard:2019bmu, JETSCAPE:2020mzn}.
Among the observables that quantify the effects of initial state fluctuations are the cumulants~\cite{Teaney:2010vd}, which can be used to decompose both the spatial and momentum anisotropies.
The cumulants of the eccentricity distribution describe the shape of the initial state fluctuations.
There are different models describing this initial state and its fluctuations.
One of the simplest analytical descriptions is the Bessel-Gaussian model (BGM)~\cite{Voloshin:2007pc} in which many sources of entropy production result in a two-dimensional Gaussian shape of fluctuations, where all higher-order ($m \geq 4$) eccentricity cumulants, $\varepsilon_2\{m\}$, are degenerate and equal to the reaction plane eccentricity
\begin{equation}
    \label{eq:BGM}
    \eptt > \eptf = \eprp\ .
\end{equation}
The BGM is only applicable to small fluctuations.
The elliptic-power model (EPM) is a more general analytical model, which is also applicable for the description of large fluctuations~\cite{Yan:2014afa}.
In the EPM, the degeneracy of higher-order cumulants is broken and therefore
\begin{equation}
    \label{eq:EPM}
    \eptt > \eptf > \eprp\ .
\end{equation}
There are a number of Monte Carlo models of the initial state that either implement a specific physical mechanism of pre-equilibrium entropy production, for example IP-Glasma~\cite{Schenke:2012wb}, or do not assume a specific mechanism, such as \TRENTo~\cite{Moreland:2014oya}.
In \TRENTo, the entropy production in the participant zone is parameterized by the transverse density of the participating nucleons, i.e. the participant thickness.
Each of these models is characterized by specific signatures of the eccentricity cumulants and their relationships.
The importance of subnucleonic degrees of freedom for modeling fluctuations in central collisions has been highlighted in Ref.~\cite{Song:2010mg}.

One can probe the initial state eccentricities by comparing the ratios of eccentricity cumulants defined by \cref{eq:eccentricity} to equivalent flow observables, if the fluid response is linear as given by \cref{eq:linear_scaling}.
These observables can be constructed through angular correlations of the emitted particles and the symmetry planes
\begin{align}
    \label{eq:v2sp}
    \vtsp{} & = \avg{\avg{\cos2(\varphi_1-\PsiSP)}}~,\\
    \label{eq:v22}
    \vtetagapnoone & =\sqrt{ \avg{\avg{\cos2(\varphi_1-\varphi_2)}}}=\avg{v^2_2}^{\frac{1}{2}}~,\\
    \label{eq:v24}
    \vtf & = \left[2\avg{\avg{\cos2(\varphi_1-\varphi_2)}}^2 - \avg{\avg{\cos2(\varphi_1-\varphi_2+\varphi_3-\varphi_4)}}\right]^{\frac{1}{4}}  = \left[2\avg{v^2_2}^2 - \avg{v^4_2}\right]^{\frac{1}{4}}\ .
\end{align}
Here $\varphi_i$ ($i=1\dots4$) are the azimuthal angles of the produced particles, $\avg{\ldots}$ is the average over all collisions, and  $\avg{\avg{\ldots}}$ is the average over all particles in a given transverse momentum window in all collisions.
The \vtsp is the elliptic flow measured relative to the spectator plane with angle \PsiSP, where fluctuations of the spectator symmetry planes are assumed to be small.
The \vtt and \vtf are estimators of elliptic flow coefficients whose equations, including \pt-dependence, are derived from the two- and four-particle cumulants~\cite{Borghini:2000sa,Bilandzic:2010jr}.
The variable $|\Delta\eta|$ in~\cref{eq:v22} refers to the minimum pseudorapidity ($\eta$) separation between the two correlated particles.
The observables \vtsp, \vtf, and \vtetagapnoone may deviate from $v_2\{\PsiRP\}$, which is the unmeasurable momentum anisotropy relative to the reaction plane.
The comparison of ratios of the measured anisotropic flow coefficients to their respective eccentricities is instrumental for identifying the most realistic model of the initial state.

The transverse momentum dependence of the flow coefficients \vn allows one to study if there is a common origin of anisotropy at different transverse momentum scales due to initial state eccentricity fluctuations.
The main source of anisotropy for soft ($\pt < \SI{3}{\GeVc}$) particle production is the collective expansion of the QGP~\cite{Qiu:2011iv}.
At significantly larger \pt, the dominant contribution to \vn is the path length dependent energy loss of the jet propagating through the azimuthally asymmetric medium oriented along \PsiPP~\cite{ALICE:2012vgf}.
A significant transverse momentum dependence of flow fluctuations relative to the participant planes is observed~\cite{ALICE:2012vgf} in central collisions and within large experimental uncertainties in midcentral collisions.
Given that the spectators only probe the early time evolution of the collision and provide sensitivity to a different symmetry plane, the transverse momentum dependence of the ratio \vtspvtf can be used as an additional observable to investigate the origin of flow fluctuations.

An accurate determination of the flow coefficients relative to the spectator plane, in particular of the elliptic flow \vtsp, is also crucial for the search of the chiral magnetic effect (CME)~\cite{STAR:2021mii, STAR:2021pwb}.
Because the magnetic field orientation is correlated more strongly to the spectator deflection than to the participant plane orientation, the measurement relative to both planes allows the separation of the CME signal from the background correlations coupled to the elliptic flow~\cite{Voloshin:2004vk,Wang:2009kd,Bzdak:2009fc, Xu:2017qfs,Voloshin:2018qsm, STAR:2021pwb}.

In this letter, the charged particle elliptic flow relative to the neutron spectator deflection, \vtsp, and to the participant plane, quantified by \vtetagap and \vtf, is reported as a function of centrality and transverse momentum for collisions of Pb ions at a center-of-mass energy per nucleon--nucleon pair of $\sqrts=\SI{2.76}{TeV}$ and Xe ions at $\sqrts=\SI{5.44}{TeV}$.
The comparison with different initial state models, such as BGM, EPM, and \TRENTo, and the scaling properties of $v_2/\varepsilon_2$ are also presented.

ALICE is a dedicated heavy-ion experiment at the LHC.
A detailed description of the detector and its performance can be found elsewhere~\cite{Collaboration_2008,ALICE:2014sbx}.
The trajectories of the charged particles are reconstructed using the Inner Tracking System (ITS) and the Time Projection Chamber (TPC)~\cite{ALICE:1999cls,Alme:2010ke}.
The neutron spectator deflection is estimated using the neutron Zero Degree Calorimeters (ZDC)~\cite{ALICE:1999edx}.
The ZDC consist of two detectors with a $2\times2$ transverse segmentation.
During the \PbPb (\XeXe) data-taking, the ZDC were placed 114 m (112.5 m) away from the collision point on both sides of the experiment, named A and C side, covering the pseudorapidity interval $|\eta|>8.78~(8.77)$.
The collision centrality is reconstructed using the energy deposition in the forward V0 scintillator arrays~\cite{ALICE:2013axi}.

About 10 (1) million of minimum bias~\cite{ALICE:2014sbx} \PbPb (\XeXe) collisions collected in the year 2010 (2017) are analyzed.
This corresponds to collisions in the 5--70\% centrality range and a reconstructed primary collision vertex located within $\pm\SI{10}{cm}$ along the beam direction from the nominal interaction point.
Outside of this centrality range, the deflection of the spectators cannot be reliably reconstructed because of the small number of detected neutron spectators in central collisions and the large fraction of spectator neutrons that are not detected by the ZDC in peripheral collisions because they are bound in charged fragments and thus deflected by the LHC magnets.

The charged particles are selected within the range of pseudorapidity $|\eta| < 0.8$ and transverse momentum $\pt > \SI{0.2}{\GeVc}$.
The tracks are reconstructed using the combined information from the ITS and the TPC detectors.
A minimum of 70 TPC space points (out of a maximum of 159) and a minimum of two ITS hits, with at least one in one of the two innermost layers, are required for all tracks.
A reduced $\chi^2$ per TPC space point (ITS hit) in the range 0.1--4 (less than 36) is required.
Only tracks with a distance of closest approach to the primary vertex position smaller than \SI{3.2}{cm} in the beam direction and within a radius of \SI{2.4}{cm} in the transverse plane are considered.

The observables in \cref{eq:v2sp,eq:v22,eq:v24}, are calculated using flow vectors $q_n^K$ of harmonic $n$ in each collision as
\begin{equation}
\label{eq:qvector}
    q_n^K = x_n^K + iy_n^K = \left. \sum_{j\in K} (w_j)^p e^{in\varphi_j} \right/ \sum_{j\in K} (w_j)^p~,
\end{equation}
where $K$ denotes either the reconstructed tracks $K=T$ or the two ZDC subdetectors $K=A,C$.

For the second harmonic flow vector of the reconstructed tracks $q_2^{T}$, the $w_j$ and $\varphi_j$ are the weight and azimuthal angle of the $j$-th track, respectively.
The weights in the $q_2^{T}$ calculation are defined by the inverse product $w=1/(\epsilon\dNdphi)$ of the \pt-dependent reconstruction efficiency $\epsilon$ and the measured non-uniform azimuthal angle distribution $\dNdphi$ of the charged particles.
The $\dNdphi$ is calculated as a function of the pseudorapidity of the particles and the primary vertex position along the beam direction.
The impact of the correction for the non-uniform azimuthal acceptance of the tracks is negligible.
The reconstruction efficiency correction is derived from a Monte Carlo simulation using the HIJING~\cite{Gyulassy:1994ew} heavy-ion event generator to simulate collisions and GEANT3~\cite{Brun:1119728} to model particle transport through the ALICE detector material.
It is applied to all measurements of the \pt-integrated flow coefficients.
The power $p$ in~\cref{eq:qvector} corresponds to the correction of the non-uniform azimuthal acceptance and reconstruction efficiency in the multi-particle cumulant method~\cite{Bilandzic:2013kga}.
It only differs from unity for $q_2^{T}$.
The \vtetagap and \vtf are calculated from flow vectors defined in~\cref{eq:qvector} using the multi-particle cumulant approach~\cite{Bilandzic:2013kga}.
In the measurement of \vtetagap, a pseudorapidity separation of $|\Delta\eta| > 1$ is applied as it largely suppresses few-particle correlations (so called non-flow), such as those originating from jets or resonance decays~\cite{ALICE:2014dwt}.

For the first harmonic flow vectors of the ZDC, $q_1^{A}$ and $q_1^{C}$, $\varphi_j$ corresponds to the angle of the center of the $j$-th calorimeter segment and $w_j$ to its measured energy deposition.
The indices $A$ and $C$ stand for the subdetectors located on either side of the interaction point.
The neutron spectator deflection is estimated with these flow vectors.
The beam configuration can result in an average shift of the spectator neutron distributions in the transverse plane of the ZDC that are corrected by a data-driven procedure (re-centering)
\begin{equation}
\label{eq:rec}
q_n' = q_n - \avg{q_n}\ .
\end{equation}
To account for changing beam conditions, the re-centering correction is calculated as a function of the centrality, and the three dimensional position of the collision vertex for each data-taking period.
The elliptic flow \vtsp is calculated from an average of three independent estimates using the scalar product method with mixed harmonics~\cite{Selyuzhenkov:2007zi}
\begin{equation}
    \label{eq:v2spe}
    \vtsp{} = \frac{2}{3}\left(\frac{\avg{x_2^{T}x_1^{A}x_1^C}}{\avg{x_1^{A}x_1^C}} - \frac{\avg{x_2^{T}y_1^{A}y_1^C}}{\avg{y_1^{A}y_1^C}} +\sqrt{\frac{\avg{y_2^{T}x_1^{A}y_1^C}\avg{y_2^{T}y_1^{A}x_1^C}}{\avg{x_1^{A}x_1^C}\avg{y_1^{A}y_1^C}}}\right)\ ,
\end{equation}
which corresponds to a measurement of elliptic flow relative to the common spectator plane $\PsiSP$.
The numerical coefficient (2/3) arises from (1/3) which corresponds to an average of the three independent terms in Eq.(11) and an additional
factor two from an expectation that correlators in the numerators of individual terms gives $(1/4) v_2\{\Psi_{\mathrm{SP}}\} R_1^2\{\mathrm{ZDC}\}$ and in the denominators $(1/2) R_1^2\{\mathrm{ZDC}\}$, where $R_1\{\mathrm{ZDC}\}$ is a convolution of the directed flow of spectators with the ZDC response.

The systematic uncertainties are evaluated from variations of the analysis procedure.
The total systematic uncertainty is given by the square root of the sum in quadrature of the differences in the results obtained from each significant variation.
A variation is considered significant, if $\Delta R/\Delta \sigma > 1$~\cite{Barlow:2002yb}, where $\Delta R$ is the difference between results from a given variation and the default analysis procedure.
The variable $\Delta\sigma$ is the square root of the difference between the corresponding squared statistical uncertainties.

The description of systematic uncertainties related to measurements of \vtetagap{} and \vtf{} is presented in Ref.~\cite{ALICE:2018rtz}.
The details of the systematic uncertainty evaluation for \vtsp are described below.
The largest source of systematic uncertainty is due to spurious correlations which are not corrected by the recentering procedure.
These affect the denominators in~\cref{eq:v2spe}, which are determined by the diagonal correlations $\avg{x^{A}_1x^{C}_1}$ and $\avg{y^{A}_1y^{C}_1}$.
The corresponding relative uncertainty, labeled as ZDC scale uncertainty, is evaluated from the off-diagonal correlations $\avg{x_1^A y_1^C}$ and $\avg{y_1^A x_1^C}$, which should not contain physical signal, divided by the diagonal correlations.
This uncertainty is found to be about $4\%$ ($9\%$) in \PbPb (\XeXe) collisions for all centrality classes and it is assigned as a fully correlated uncertainty for \vtsp results.
The observed negative sign of the diagonal correlations indicates a deflection of the neutron spectators in, on average, opposite directions~\cite{ALICE:2013xri}.
The off-diagonal correlations may be non-zero because of the specific beam conditions during the LHC operations that may affect the distribution of neutrons impinging ZDC, such as the beam divergence at the interaction point.

The second largest contribution to the relative uncertainty comes from the differences among the three independent estimates of \vtsp corresponding to the three terms in~\cref{eq:v2spe} and amounts to lower than 2\%.
The uncertainty due to the implementation of the flow vector re-centering procedure, defined by~\cref{eq:rec}, is evaluated by comparing to results from an iterative procedure.
In the latter, each of the four iterations consists of one coarse-binned four-dimensional re-centering step followed by four fine-grained one-dimensional steps as a function of the centrality and the three dimensional position of the collision vertex.
The corresponding relative variation is below 1\%.
Introducing an additional equalization of the average ZDC signals in the flow vector calculation in~\cref{eq:qvector} amounts to a relative uncertainty less than $1\%$.

The systematic uncertainties related to the reconstruction of charged particle tracks and to the secondary particle contamination are evaluated by varying the track selection criteria.
A relative uncertainty from changing the selection of the distance to the primary vertex of the collision as a function of particle \pt, the minimum number of TPC clusters and the reduced $\chi^2$ per TPC cluster is about 1.5\%.
No significant difference between results for positively and negatively charged particles is observed.
A relative uncertainty of 0.7\% on the \pt-dependent track reconstruction efficiency was estimated from the difference between the results obtained with an efficiency calculation using all primary tracks and those using only tracks of primary $\pi$, K, p, $\mu$, and e.
The uncertainty due to the non-uniform acceptance correction, estimated from the difference of results with and without applying the $1/(\dNdphi)$ weights in~\cref{eq:qvector}, is not significant.
An uncertainty due to the centrality determination reaching up to 2\% in the most peripheral collisions is estimated from the difference of the results obtained when using an alternative centrality estimator based on the multiplicity of hits in the ITS.
The effects of the TPC misalignment and the beam optics at the interaction point are estimated to be smaller than 0.5\% by comparing results obtained with different magnetic field polarities of the ALICE solenoid.
The impact on the results of the variation of charged particle acceptance in pseudorapidity within the fiducial volume of ALICE is probed by reducing the accepted range of the position of the reconstructed primary vertex of the collisions along the beam direction from the nominal $\pm\SI{10}{cm}$ to $\pm\SI{8}{cm}$.
It yields no significant change of \vtsp.
For the ratios $\vtetagap/\vtf$ and \vtspvtf most of the common sources of systematic uncertainties cancel out.
In addition, the bin-to-bin uncorrelated uncertainties cancel in the \pt-dependence of these ratios.
\begin{figure}[t]
\centering
\includegraphics[width=0.95\textwidth]{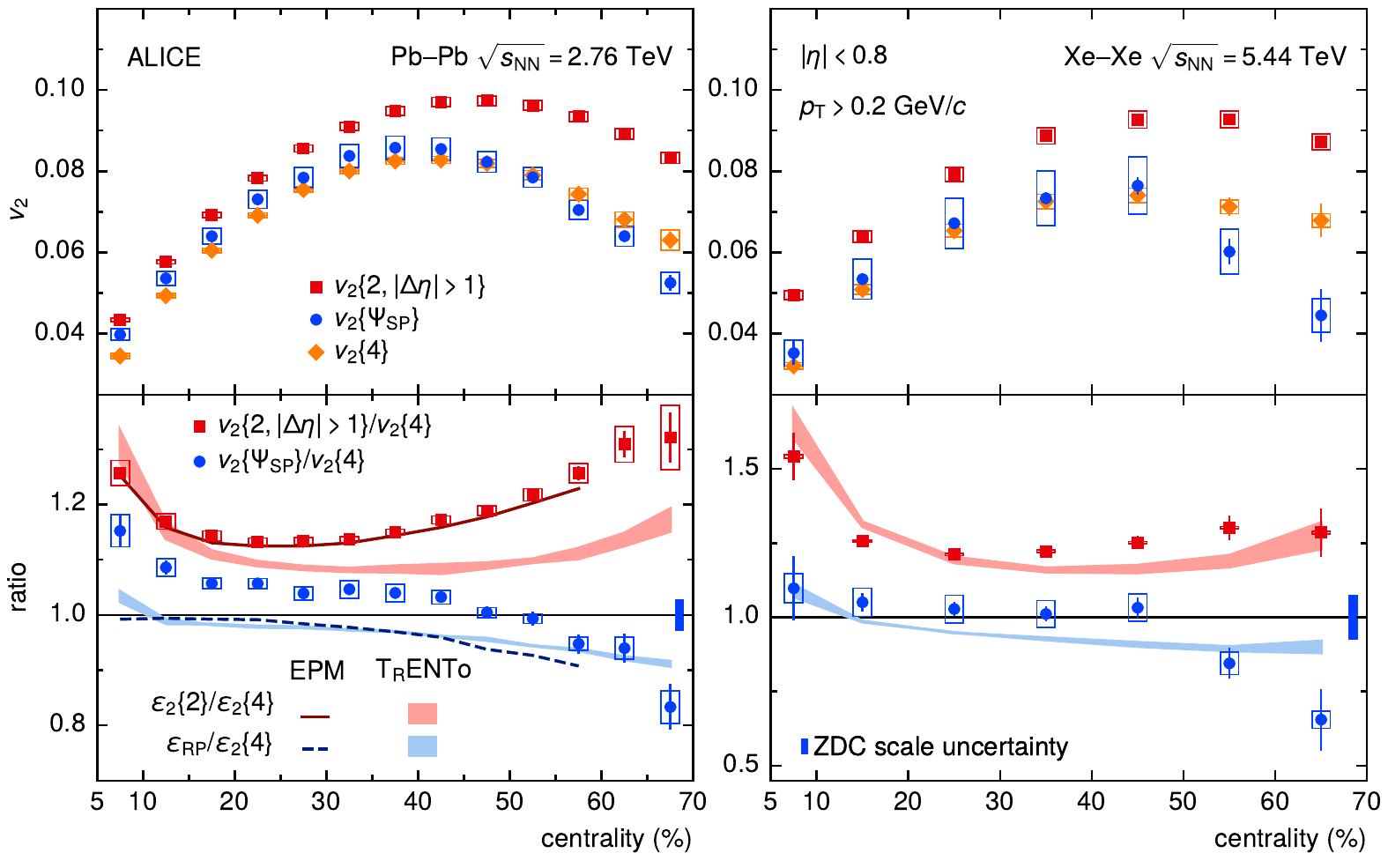}
\caption{
(color online)
(upper panels) Elliptic flow relative to the spectator plane, \vtsp, and to the participant plane, \vtetagap and \vtf, as a function of centrality in \PbPb (left) and \XeXe (right) collisions.
(bottom panels) Ratios of the elliptic flow \vtsp and \vtetagap to \vtf.
The dashed (solid) lines show the eccentricity ratios of \eprp (\eptt) to \eptf from the elliptic power model.
The corresponding eccentricity ratios for \TRENTo are shown as solid bands.
The error bars (open boxes) indicate statistical (systematic) uncertainties.
The bin-to-bin uncorrelated uncertainties and the correlated ones are combined for \vtsp results.
For the ratio \vtspvtf, the ZDC scale uncertainties are shown separately as solid boxes centered at unity on the right side of the lower panels.
}
\label{fig:v2_cent}
\end{figure}

The results for the elliptic flow coefficients relative to the spectator symmetry plane, \vtsp{}, and the participant plane, \vtt and \vtf, as a function of the centrality in \PbPb and \XeXe collisions are shown in the upper panels of~\cref{fig:v2_cent}.
In both \PbPb and \XeXe collisions, the maximum of \vtsp is located in the 35--45\% centrality range with a hint of being shifted towards more peripheral collisions for smaller systems.
The difference between \vtetagap and \vtf is due to fluctuations in the participant zone, which have been extensively studied~\cite{ALICE:2011ab,STAR:2011ert,PHOBOS:2010ekr,PHOBOS:2007wlf}.
In peripheral collisions, this difference may also originate from residual few-particle correlations.
In midcentral collisions, \vtsp and \vtf are similar as expected from the BGM (see \cref{eq:BGM}).

To study the differences between the flow observables in detail, the lower panels of ~\cref{fig:v2_cent} show the centrality dependence of the ratios of \vtetagap and \vtsp to \vtf as well as the ratios of the respective eccentricity cumulants calculated from the EPM and the \TRENTo models.
The width of the bands, which represent the \TRENTo calculations, in~\cref{fig:v2_cent} indicate the differences between two configurations that are tuned to different experimental data.
The model in Ref.~\cite{Bernhard:2019bmu} uses \PbPb collisions at $\sqrts=\SI{2.76}{TeV}$ and $\sqrts=\SI{5.02}{TeV}$, whereas Ref.~\cite{JETSCAPE:2020mzn} uses data of Au--Au collisions at $\sqrts=\SI{0.2}{TeV}$ and \PbPb collisions at $\sqrts=\SI{2.76}{TeV}$.
The deformed Woods-Saxon parametrization of Xe nuclei is adopted from Ref.~\cite{ALICE:2018cpu}.
The considered model calculations do not take into account the subnucleonic degrees of freedom.
It was demonstrated in Ref.~\cite{ALICE:2018lao} based on two-particle cumulant measurements that at the LHC they only impact very central (0--5\%) \PbPb collisions, which is outside of the reported centrality range.
The EPM model parameters, which relies on significant differences between higher order cumulants, are only accessible with the larger data sample of \PbPb collisions at $\sqrts=\SI{5.02}{TeV}$~\cite{ALICE:2018rtz}.

In both central and peripheral \PbPb collisions, the ratio \vtspvtf strongly deviates from unity, which is the value expected from the BGM for the eccentricity ratio.
In midcentral collisions, the measured ratio is smaller than in central collisions, but it is significantly larger than the BGM (unity), EPM and \TRENTo model calculations.
The later indicates that the spectator plane becomes decorrelated from the reaction plane and simultaneously more correlated with the participant plane.
Thus \vtsp becomes closer to \vtetagap, which makes the ratio \vtspvtf above unity.
Similar trends are observed in \XeXe collisions, however due to the limited size of the data sample the deviations from unity are not significant within the uncertainties except in the most peripheral collisions.
A decorrelation in peripheral collisions is expected due to the small number of sources of particle production (number of participants), as well as due to the decreasing fraction of energy carried by the neutrons measured with the ZDC relative to the unmeasured energy of protons and other charged nuclear fragments.

In central collisions, despite the largest number of particle-producing sources, the fluctuations are also large and deviations of the \vtspvtf ratio from unity are growing, which is in contrast to BGM expectations.
The deviation from unity is also large compared to the EPM calculations, which yield a value close to unity in central collisions.
This can be either due to (a) a small number of spectators emitted in these collisions which increases the spectator plane fluctuations (decorrelation) around the reaction plane and consequently increases the magnitude of \vtsp, (b) the strengthening of the correlation between \PsiSP and $\Psi_2$ angles, or (c) specific geometry fluctuations and/or correlated local particle production in individual participating nucleon--nucleon interactions that reduce \vtf.
In midcentral and peripheral collisions, both models predict a small deviation from unity for $\eprp/\eptf$, which is expected for small fluctuations given by the second equality of \cref{eq:BGM} and due to the small skewness observed in Ref.~\cite{ALICE:2018rtz}.
The ratio of $\eptt/\eptf$ from the \TRENTo model shows smaller fluctuations compared to the tuned-on-data EPM calculations and the measured $\vtetagap/\vtf$.

\begin{figure}[t]
\centering
\includegraphics[width=1.0\textwidth]{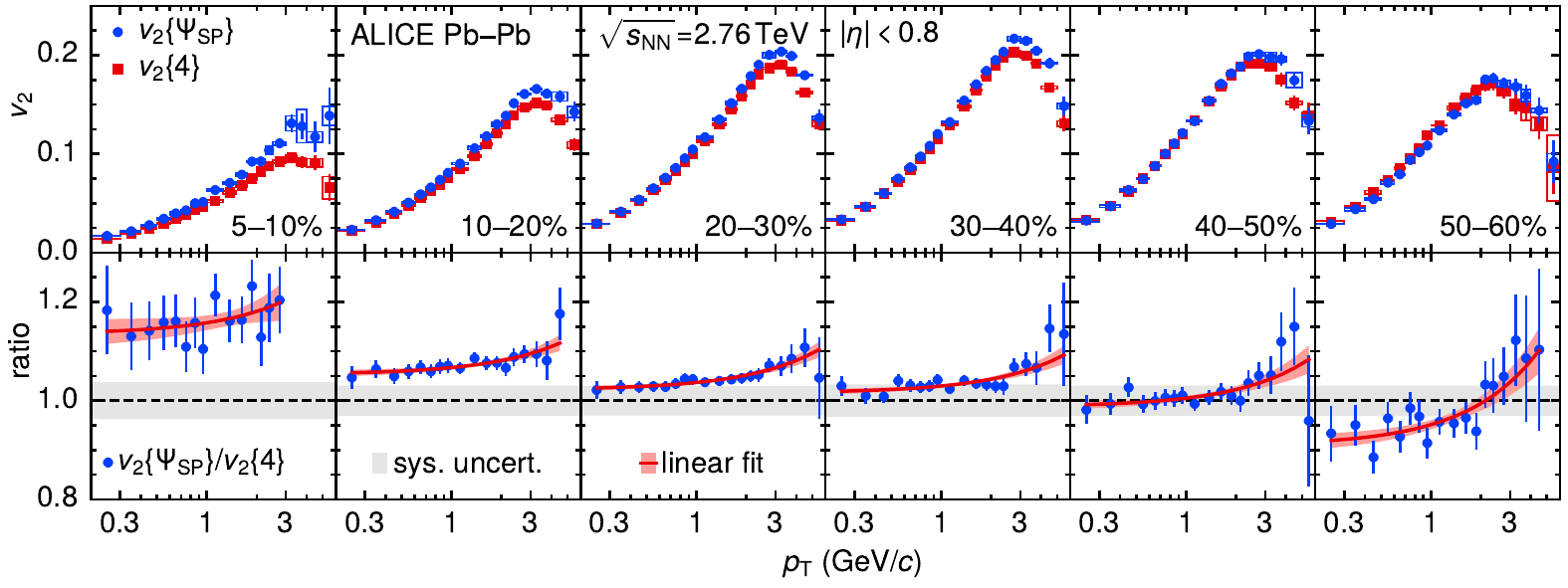}
\caption{
(color online)
(upper panels) Elliptic flow relative to the spectator plane, \vtsp, and to the participant plane, \vtf, as a function of transverse momentum in different centrality classes for \PbPb collisions.
The linear fits to the ratio \vtspvtf are shown as red lines for the individual centrality classes.
The error bars (open boxes) indicate statistical (systematic) uncertainties.
The correlated uncertainties, related to the ZDC, and bin-to-bin uncorrelated ones are combined for \vtsp (upper panels).
For the ratio \vtspvtf (lower panels), the correlated uncertainty is shown by the grey band at unity.
}
\label{fig:v2_pt}
\end{figure}

The upper panels of~\cref{fig:v2_pt} show the dependence of the elliptic flow coefficients \vtsp and \vtf on transverse momentum for different centrality classes in \PbPb collisions.
The lower panels of~\cref{fig:v2_pt} show the ratio \vtspvtf as a function of \pt for different centrality classes.
To quantify the possible dependence of \vtspvtf on \pt, the ratio was fitted with a linear function in the range \SIrange{0.2}{6}{\GeVc}.
The slopes of the linear fit in different centrality classes are:
$2.31 \pm 2.17$ (5--10\%),
$1.41 \pm 0.48$ (10--20\%),
$1.48 \pm 0.36$ (20--30\%),
$1.40 \pm 0.42$ (30--40\%),
$1.74 \pm 0.65$ (40--50\%),
$4.29 \pm 1.34$ (50--60\%) in units of $100/(\mathrm{GeV}/c)$.
There are indications for a non-flat transverse momentum dependence of \vtspvtf with a statistical significance of about 2.5--4$\sigma$ in all centrality classes except for 5--10\% central collisions where it is only 1$\sigma$.
These observations are important for understanding to which extent the momentum anisotropy at low and intermediate transverse momentum has a common origin from the fluctuating initial state anisotropy in the coordinate space and quantifying the effects of momentum-dependent flow magnitude and angle decorrelation~\cite{ALICE:2017lyf, CMS:2015xmx}.

\begin{figure}[t]
\centering
\includegraphics[width=0.69\textwidth]{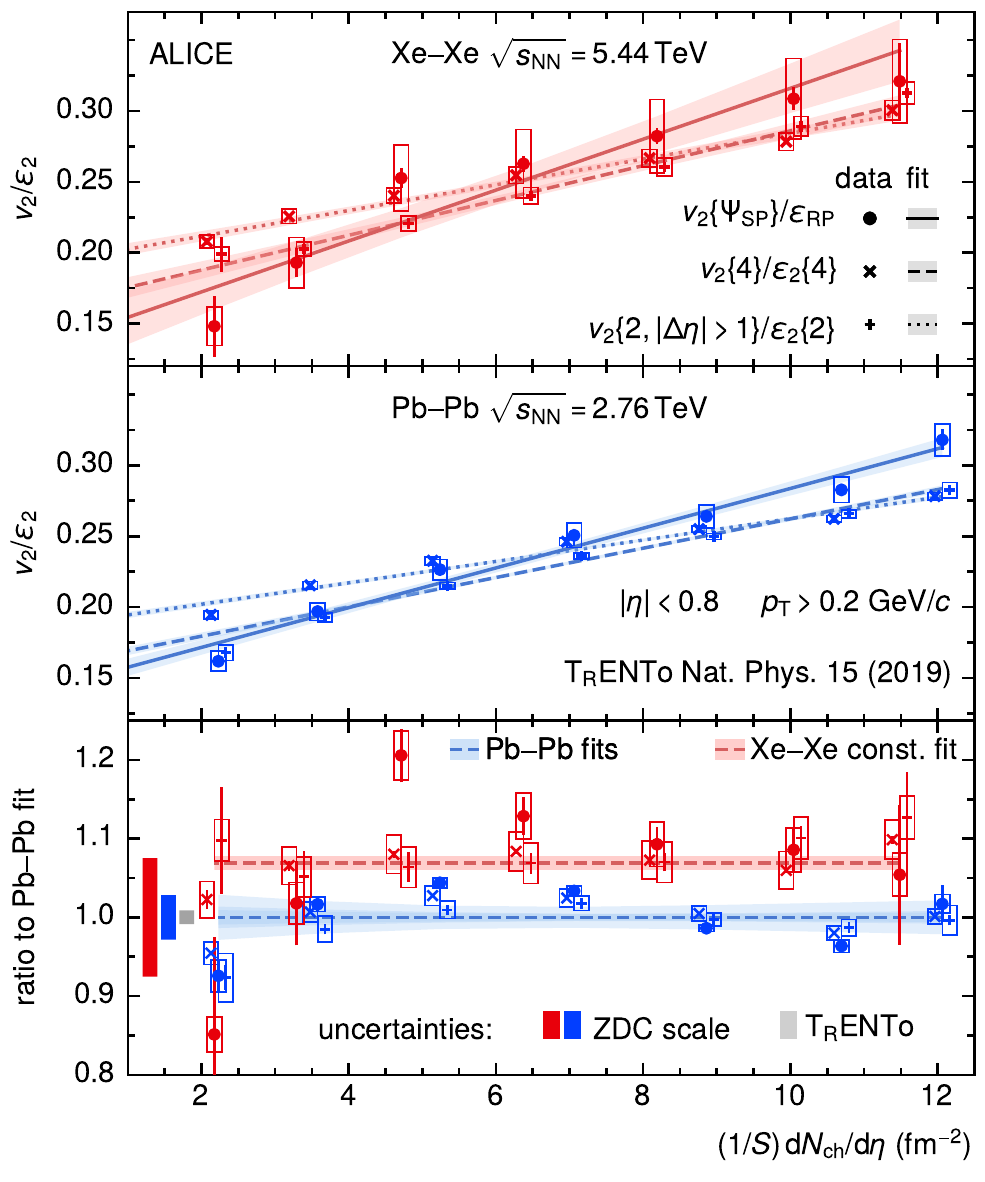}
\caption{
(color online)
Ratios of $v_2/\varepsilon_2$ as a function of $(1/S)\dNchdEta$ in \XeXe (top panel) and \PbPb (middle panel) collisions.
The Linear fit functions to $v_2/\varepsilon_2$ in the top and middle panels are shown by lines.
The bands show the uncertainty of the fit.
(bottom panel) The ratio of $v_2/\varepsilon_2$ to the linear \PbPb fits shown in the middle panel.
The data points of \vtetagap, \vtf are shifted by $-0.1$ and $+0.1$ along $(1/S)\dNchdEta$ for better visibility.
The error bars (open boxes) indicate statistical (systematic) uncertainties.
The ZDC scale uncertainty, due to residual correlations in the determination of the denominators in~\cref{eq:v2spe}, and the bin-to-bin uncorrelated one are combined for \vtsp (top and middle panels).
For the ratio of \vtsp to the \PbPb fit (bottom panel), the ZDC scale uncertainty is separated and is shown by the solid boxes on the left side of the bottom panel.
The grey box at unity in the bottom panel on the left side shows the variation depending on the \TRENTo configuration.
The blue horizontal band (bottom panel) represents the relative uncertainties of the individual linear fits to the \PbPb data from the middle panel.
The red horizontal line (bottom panel) shows a combined fit of a constant function to the \XeXe data and its uncertainty.
}
\label{fig:v2_multscaling}
\end{figure}

The scaling behavior of the elliptic flow with the initial state eccentricities, as suggested by~\cref{eq:mult_scaling}, is shown in~\cref{fig:v2_multscaling}.
The top and middle panels present the ratio $v_2/\varepsilon_2$ as a function of the charged particle multiplicity density per unit of the nuclei overlap area $(1/S)\dNchdEta$ for the elliptic flow coefficients \vtetagap, \vtf, and \vtsp.
The area $S$ and the corresponding eccentricities are calculated from the \TRENTo model using the configuration from Ref.~\cite{Bernhard:2019bmu}.
The experimental data points for $\dNchdEta$ as a function of centrality are taken from Ref.~\cite{ALICE:2010mlf} for \PbPb and from Ref.~\cite{ALICE:2018cpu} for \XeXe collisions.
The lines represent linear fits to the ratios $\vtsp/\eprp$, $\vtetagap/\eptt$, and $\vtf/\eptf$.

Deviations from the scaling behavior are quantified in the bottom panel of~\cref{fig:v2_multscaling}, which shows $v_2/\varepsilon_2$ divided by the respective linear fits of the \PbPb measurement.
The small differences between eccentricities calculated from the two \TRENTo configurations are reflected in the height of the solid gray box on the left side.
The splitting between the individual observables for a given collision system may be sensitive to the chosen association of eccentricities to flow coefficients.
A relative deviation of $(7.0 \pm 0.9)\%$ from the scaling given by~\cref{eq:mult_scaling} between the two collision systems is extracted from a constant line fit to the \XeXe data and by propagating uncertainties of the \PbPb data.
An additional variation of $+1.8\%$ is estimated from the difference between the two \TRENTo configurations.
It is similar for the different flow coefficients and only weakly depends on the multiplicity density or collision centrality.
This collision system dependence may be sensitive to the details of the initial state models~\cite{Voloshin:2007af,ALICE:2018lao} and to viscous effects in the expansion of the QGP~\cite{Song:2008si}.
Together with the results on the centrality and transverse momentum dependence, shown in \cref{fig:v2_cent,,fig:v2_pt}, the scaling presented in \cref{fig:v2_multscaling} provides new experimental constraints to both the initial conditions and the transport coefficients of the QGP produced in heavy-ion collisions.

In summary, the elliptic flow of charged particles at midrapidity ($|\eta|<0.8$) relative to the neutron spectator plane \vtsp is measured in \PbPb at \Epbpb and \XeXe at \Exexe collisions as a function of the centrality and as a function of the transverse momentum in different centrality classes.
The measurements are compared to results of elliptic flow relative to the participant plane obtained from two and four-particle correlations.
Both in central and peripheral collisions, the ratio \vtspvtf strongly deviates from the respective ratio of the eccentricities, $\eprp/\eptf$, predicted by the \TRENTo and EPM models of the initial state fluctuations.
The observations can be interpreted as a decorrelation of the neutron spectator plane and the reaction plane due to the fragmentation of the recoil from the colliding nuclei which indicates an incomplete description of initial state fluctuations in the models.
A transverse momentum dependence of the ratio \vtspvtf is observed with 2.5--4$\sigma$ significance in all but the most central collisions.
This is an important experimental input for understanding whether the momentum anisotropies at low and intermediate transverse momentum have a common origin in the initial state fluctuations.
The ratios of elliptic flow coefficients to the initial state eccentricities for \XeXe and \PbPb collisions reveal a relative deviation from the scaling with the initial entropy density of $(7.0 \pm 0.9)\%$ with an additional variation of $+1.8\%$ when including RHIC data in the \TRENTo parameter extraction.
The deviation is independent of the charged particle multiplicity density, which provides new experimental constraints for viscous effects in the expansion of the quark--gluon plasma produced in heavy-ion collisions at the LHC.

\newenvironment{acknowledgement}{\relax}{\relax}
\begin{acknowledgement}
\section*{Acknowledgements}

The ALICE Collaboration would like to thank all its engineers and technicians for their invaluable contributions to the construction of the experiment and the CERN accelerator teams for the outstanding performance of the LHC complex.
The ALICE Collaboration gratefully acknowledges the resources and support provided by all Grid centres and the Worldwide LHC Computing Grid (WLCG) collaboration.
The ALICE Collaboration acknowledges the following funding agencies for their support in building and running the ALICE detector:
A. I. Alikhanyan National Science Laboratory (Yerevan Physics Institute) Foundation (ANSL), State Committee of Science and World Federation of Scientists (WFS), Armenia;
Austrian Academy of Sciences, Austrian Science Fund (FWF): [M 2467-N36] and Nationalstiftung f\"{u}r Forschung, Technologie und Entwicklung, Austria;
Ministry of Communications and High Technologies, National Nuclear Research Center, Azerbaijan;
Conselho Nacional de Desenvolvimento Cient\'{\i}fico e Tecnol\'{o}gico (CNPq), Financiadora de Estudos e Projetos (Finep), Funda\c{c}\~{a}o de Amparo \`{a} Pesquisa do Estado de S\~{a}o Paulo (FAPESP) and Universidade Federal do Rio Grande do Sul (UFRGS), Brazil;
Bulgarian Ministry of Education and Science, within the National Roadmap for Research Infrastructures 2020-2027 (object CERN), Bulgaria;
Ministry of Education of China (MOEC) , Ministry of Science \& Technology of China (MSTC) and National Natural Science Foundation of China (NSFC), China;
Ministry of Science and Education and Croatian Science Foundation, Croatia;
Centro de Aplicaciones Tecnol\'{o}gicas y Desarrollo Nuclear (CEADEN), Cubaenerg\'{\i}a, Cuba;
Ministry of Education, Youth and Sports of the Czech Republic, Czech Republic;
The Danish Council for Independent Research | Natural Sciences, the VILLUM FONDEN and Danish National Research Foundation (DNRF), Denmark;
Helsinki Institute of Physics (HIP), Finland;
Commissariat \`{a} l'Energie Atomique (CEA) and Institut National de Physique Nucl\'{e}aire et de Physique des Particules (IN2P3) and Centre National de la Recherche Scientifique (CNRS), France;
Bundesministerium f\"{u}r Bildung und Forschung (BMBF) and GSI Helmholtzzentrum f\"{u}r Schwerionenforschung GmbH, Germany;
General Secretariat for Research and Technology, Ministry of Education, Research and Religions, Greece;
National Research, Development and Innovation Office, Hungary;
Department of Atomic Energy Government of India (DAE), Department of Science and Technology, Government of India (DST), University Grants Commission, Government of India (UGC) and Council of Scientific and Industrial Research (CSIR), India;
National Research and Innovation Agency - BRIN, Indonesia;
Istituto Nazionale di Fisica Nucleare (INFN), Italy;
Japanese Ministry of Education, Culture, Sports, Science and Technology (MEXT) and Japan Society for the Promotion of Science (JSPS) KAKENHI, Japan;
Consejo Nacional de Ciencia (CONACYT) y Tecnolog\'{i}a, through Fondo de Cooperaci\'{o}n Internacional en Ciencia y Tecnolog\'{i}a (FONCICYT) and Direcci\'{o}n General de Asuntos del Personal Academico (DGAPA), Mexico;
Nederlandse Organisatie voor Wetenschappelijk Onderzoek (NWO), Netherlands;
The Research Council of Norway, Norway;
Commission on Science and Technology for Sustainable Development in the South (COMSATS), Pakistan;
Pontificia Universidad Cat\'{o}lica del Per\'{u}, Peru;
Ministry of Education and Science, National Science Centre and WUT ID-UB, Poland;
Korea Institute of Science and Technology Information and National Research Foundation of Korea (NRF), Republic of Korea;
Ministry of Education and Scientific Research, Institute of Atomic Physics, Ministry of Research and Innovation and Institute of Atomic Physics and University Politehnica of Bucharest, Romania;
Ministry of Education, Science, Research and Sport of the Slovak Republic, Slovakia;
National Research Foundation of South Africa, South Africa;
Swedish Research Council (VR) and Knut \& Alice Wallenberg Foundation (KAW), Sweden;
European Organization for Nuclear Research, Switzerland;
Suranaree University of Technology (SUT), National Science and Technology Development Agency (NSTDA), Thailand Science Research and Innovation (TSRI) and National Science, Research and Innovation Fund (NSRF), Thailand;
Turkish Energy, Nuclear and Mineral Research Agency (TENMAK), Turkey;
National Academy of  Sciences of Ukraine, Ukraine;
Science and Technology Facilities Council (STFC), United Kingdom;
National Science Foundation of the United States of America (NSF) and United States Department of Energy, Office of Nuclear Physics (DOE NP), United States of America.
In addition, individual groups or members have received support from:
Marie Sk\l{}odowska Curie, Strong 2020 - Horizon 2020, European Research Council (grant nos. 824093, 896850, 950692), European Union;
Academy of Finland (Center of Excellence in Quark Matter) (grant nos. 346327, 346328), Finland;
Programa de Apoyos para la Superaci\'{o}n del Personal Acad\'{e}mico, UNAM, Mexico.

\end{acknowledgement}

\bibliographystyle{utphys}   
\bibliography{references.bib}

\newpage
\appendix

%
%

\section{The ALICE Collaboration}
\label{app:collab}
\begin{flushleft} 
\small

S.~Acharya\,\orcidlink{0000-0002-9213-5329}\,$^{\rm 123,131}$, 
D.~Adamov\'{a}\,\orcidlink{0000-0002-0504-7428}\,$^{\rm 85}$, 
A.~Adler$^{\rm 69}$, 
G.~Aglieri Rinella\,\orcidlink{0000-0002-9611-3696}\,$^{\rm 32}$, 
M.~Agnello\,\orcidlink{0000-0002-0760-5075}\,$^{\rm 29}$, 
N.~Agrawal\,\orcidlink{0000-0003-0348-9836}\,$^{\rm 50}$, 
Z.~Ahammed\,\orcidlink{0000-0001-5241-7412}\,$^{\rm 131}$, 
S.~Ahmad\,\orcidlink{0000-0003-0497-5705}\,$^{\rm 15}$, 
S.U.~Ahn\,\orcidlink{0000-0001-8847-489X}\,$^{\rm 70}$, 
I.~Ahuja\,\orcidlink{0000-0002-4417-1392}\,$^{\rm 37}$
% ...

\section*{Affiliation Notes}

$^{\rm I}$ Deceased\\
$^{\rm II}$ Also at: Max-Planck-Institut f\"{u}r Physik, Munich, Germany\\
$^{\rm III}$ Also at: Italian National Agency for New Technologies, Energy and Sustainable Economic Development (ENEA), Bologna, Italy\\
$^{\rm IV}$ Also at: Dipartimento DET del Politecnico di Torino, Turin, Italy\\
$^{\rm V}$ Also at: Department of Applied Physics, Aligarh Muslim University, Aligarh, India\\
$^{\rm VI}$ Also at: Institute of Theoretical Physics, University of Wroclaw, Poland\\
$^{\rm VII}$ Also at: An institution covered by a cooperation agreement with CERN\\

\section*{Collaboration Institutes}

$^{1}$ A.I. Alikhanyan National Science Laboratory (Yerevan Physics Institute) Foundation, Yerevan, Armenia\\
$^{2}$ AGH University of Krakow, Cracow, Poland\\
$^{3}$ Bogolyubov Institute for Theoretical Physics, National Academy of Sciences of Ukraine, Kiev, Ukraine\\
$^{4}$ Bose Institute, Department of Physics  and Centre for Astroparticle Physics and Space Science (CAPSS), Kolkata, India\\
$^{5}$ California Polytechnic State University, San Luis Obispo, California, United States\\
$^{6}$ Central China Normal University, Wuhan, China\\
$^{7}$ Centro de Aplicaciones Tecnol\'{o}gicas y Desarrollo Nuclear (CEADEN), Havana, Cuba\\
$^{8}$ Centro de Investigaci\'{o}n y de Estudios Avanzados (CINVESTAV), Mexico City and M\'{e}rida, Mexico\\
$^{9}$ Chicago State University, Chicago, Illinois, United States\\
$^{10}$ China Institute of Atomic Energy, Beijing, China\\
$^{11}$ Chungbuk National University, Cheongju, Republic of Korea\\
$^{12}$ Comenius University Bratislava, Faculty of Mathematics, Physics and Informatics, Bratislava, Slovak Republic\\
$^{13}$ COMSATS University Islamabad, Islamabad, Pakistan\\
$^{14}$ Creighton University, Omaha, Nebraska, United States\\
$^{15}$ Department of Physics, Aligarh Muslim University, Aligarh, India\\
$^{16}$ Department of Physics, Pusan National University, Pusan, Republic of Korea\\
$^{17}$ Department of Physics, Sejong University, Seoul, Republic of Korea\\
$^{18}$ Department of Physics, University of California, Berkeley, California, United States\\
$^{19}$ Department of Physics, University of Oslo, Oslo, Norway\\
$^{20}$ Department of Physics and Technology, University of Bergen, Bergen, Norway\\
$^{21}$ Dipartimento di Fisica, Universit\`{a} di Pavia, Pavia, Italy\\
$^{22}$ Dipartimento di Fisica dell'Universit\`{a} and Sezione INFN, Cagliari, Italy\\
$^{23}$ Dipartimento di Fisica dell'Universit\`{a} and Sezione INFN, Trieste, Italy\\
$^{24}$ Dipartimento di Fisica dell'Universit\`{a} and Sezione INFN, Turin, Italy\\
$^{25}$ Dipartimento di Fisica e Astronomia dell'Universit\`{a} and Sezione INFN, Bologna, Italy\\
$^{26}$ Dipartimento di Fisica e Astronomia dell'Universit\`{a} and Sezione INFN, Catania, Italy\\
$^{27}$ Dipartimento di Fisica e Astronomia dell'Universit\`{a} and Sezione INFN, Padova, Italy\\
$^{28}$ Dipartimento di Fisica `E.R.~Caianiello' dell'Universit\`{a} and Gruppo Collegato INFN, Salerno, Italy\\
$^{29}$ Dipartimento DISAT del Politecnico and Sezione INFN, Turin, Italy\\
$^{30}$ Dipartimento di Scienze MIFT, Universit\`{a} di Messina, Messina, Italy\\
$^{31}$ Dipartimento Interateneo di Fisica `M.~Merlin' and Sezione INFN, Bari, Italy\\
$^{32}$ European Organization for Nuclear Research (CERN), Geneva, Switzerland\\
$^{33}$ Faculty of Electrical Engineering, Mechanical Engineering and Naval Architecture, University of Split, Split, Croatia\\
$^{34}$ Faculty of Engineering and Science, Western Norway University of Applied Sciences, Bergen, Norway\\
$^{35}$ Faculty of Nuclear Sciences and Physical Engineering, Czech Technical University in Prague, Prague, Czech Republic\\
$^{36}$ Faculty of Physics, Sofia University, Sofia, Bulgaria\\
$^{37}$ Faculty of Science, P.J.~\v{S}af\'{a}rik University, Ko\v{s}ice, Slovak Republic\\
$^{38}$ Frankfurt Institute for Advanced Studies, Johann Wolfgang Goethe-Universit\"{a}t Frankfurt, Frankfurt, Germany\\
$^{39}$ Fudan University, Shanghai, China\\
$^{40}$ Gangneung-Wonju National University, Gangneung, Republic of Korea\\
$^{41}$ Gauhati University, Department of Physics, Guwahati, India\\
$^{42}$ Helmholtz-Institut f\"{u}r Strahlen- und Kernphysik, Rheinische Friedrich-Wilhelms-Universit\"{a}t Bonn, Bonn, Germany\\
$^{43}$ Helsinki Institute of Physics (HIP), Helsinki, Finland\\
$^{44}$ High Energy Physics Group,  Universidad Aut\'{o}noma de Puebla, Puebla, Mexico\\
$^{45}$ Horia Hulubei National Institute of Physics and Nuclear Engineering, Bucharest, Romania\\
$^{46}$ Indian Institute of Technology Bombay (IIT), Mumbai, India\\
$^{47}$ Indian Institute of Technology Indore, Indore, India\\
$^{48}$ INFN, Laboratori Nazionali di Frascati, Frascati, Italy\\
$^{49}$ INFN, Sezione di Bari, Bari, Italy\\
$^{50}$ INFN, Sezione di Bologna, Bologna, Italy\\
$^{51}$ INFN, Sezione di Cagliari, Cagliari, Italy\\
$^{52}$ INFN, Sezione di Catania, Catania, Italy\\
$^{53}$ INFN, Sezione di Padova, Padova, Italy\\
$^{54}$ INFN, Sezione di Pavia, Pavia, Italy\\
$^{55}$ INFN, Sezione di Torino, Turin, Italy\\
$^{56}$ INFN, Sezione di Trieste, Trieste, Italy\\
$^{57}$ Inha University, Incheon, Republic of Korea\\
$^{58}$ Institute for Gravitational and Subatomic Physics (GRASP), Utrecht University/Nikhef, Utrecht, Netherlands\\
$^{59}$ Institute of Experimental Physics, Slovak Academy of Sciences, Ko\v{s}ice, Slovak Republic\\
$^{60}$ Institute of Physics, Homi Bhabha National Institute, Bhubaneswar, India\\
$^{61}$ Institute of Physics of the Czech Academy of Sciences, Prague, Czech Republic\\
$^{62}$ Institute of Space Science (ISS), Bucharest, Romania\\
$^{63}$ Institut f\"{u}r Kernphysik, Johann Wolfgang Goethe-Universit\"{a}t Frankfurt, Frankfurt, Germany\\
$^{64}$ Instituto de Ciencias Nucleares, Universidad Nacional Aut\'{o}noma de M\'{e}xico, Mexico City, Mexico\\
$^{65}$ Instituto de F\'{i}sica, Universidade Federal do Rio Grande do Sul (UFRGS), Porto Alegre, Brazil\\
$^{66}$ Instituto de F\'{\i}sica, Universidad Nacional Aut\'{o}noma de M\'{e}xico, Mexico City, Mexico\\
$^{67}$ iThemba LABS, National Research Foundation, Somerset West, South Africa\\
$^{68}$ Jeonbuk National University, Jeonju, Republic of Korea\\
$^{69}$ Johann-Wolfgang-Goethe Universit\"{a}t Frankfurt Institut f\"{u}r Informatik, Fachbereich Informatik und Mathematik, Frankfurt, Germany\\
$^{70}$ Korea Institute of Science and Technology Information, Daejeon, Republic of Korea\\
$^{71}$ KTO Karatay University, Konya, Turkey\\
$^{72}$ Laboratoire de Physique Subatomique et de Cosmologie, Universit\'{e} Grenoble-Alpes, CNRS-IN2P3, Grenoble, France\\
$^{73}$ Lawrence Berkeley National Laboratory, Berkeley, California, United States\\
$^{74}$ Lund University Department of Physics, Division of Particle Physics, Lund, Sweden\\
$^{75}$ Nagasaki Institute of Applied Science, Nagasaki, Japan\\
$^{76}$ Nara Women{'}s University (NWU), Nara, Japan\\
$^{77}$ National and Kapodistrian University of Athens, School of Science, Department of Physics , Athens, Greece\\
$^{78}$ National Centre for Nuclear Research, Warsaw, Poland\\
$^{79}$ National Institute of Science Education and Research, Homi Bhabha National Institute, Jatni, India\\
$^{80}$ National Nuclear Research Center, Baku, Azerbaijan\\
$^{81}$ National Research and Innovation Agency - BRIN, Jakarta, Indonesia\\
$^{82}$ Niels Bohr Institute, University of Copenhagen, Copenhagen, Denmark\\
$^{83}$ Nikhef, National institute for subatomic physics, Amsterdam, Netherlands\\
$^{84}$ Nuclear Physics Group, STFC Daresbury Laboratory, Daresbury, United Kingdom\\
$^{85}$ Nuclear Physics Institute of the Czech Academy of Sciences, Husinec-\v{R}e\v{z}, Czech Republic\\
$^{86}$ Oak Ridge National Laboratory, Oak Ridge, Tennessee, United States\\
$^{87}$ Ohio State University, Columbus, Ohio, United States\\
$^{88}$ Physics department, Faculty of science, University of Zagreb, Zagreb, Croatia\\
$^{89}$ Physics Department, Panjab University, Chandigarh, India\\
$^{90}$ Physics Department, University of Jammu, Jammu, India\\
$^{91}$ Physics Department, University of Rajasthan, Jaipur, India\\
$^{92}$ Physics Program and International Institute for Sustainability with Knotted Chiral Meta Matter (SKCM2), Hiroshima University, Hiroshima, Japan\\
$^{93}$ Physikalisches Institut, Eberhard-Karls-Universit\"{a}t T\"{u}bingen, T\"{u}bingen, Germany\\
$^{94}$ Physikalisches Institut, Ruprecht-Karls-Universit\"{a}t Heidelberg, Heidelberg, Germany\\
$^{95}$ Physik Department, Technische Universit\"{a}t M\"{u}nchen, Munich, Germany\\
$^{96}$ Politecnico di Bari and Sezione INFN, Bari, Italy\\
$^{97}$ Research Division and ExtreMe Matter Institute EMMI, GSI Helmholtzzentrum f\"ur Schwerionenforschung GmbH, Darmstadt, Germany\\
$^{98}$ Saha Institute of Nuclear Physics, Homi Bhabha National Institute, Kolkata, India\\
$^{99}$ School of Physics and Astronomy, University of Birmingham, Birmingham, United Kingdom\\
$^{100}$ Secci\'{o}n F\'{\i}sica, Departamento de Ciencias, Pontificia Universidad Cat\'{o}lica del Per\'{u}, Lima, Peru\\
$^{101}$ Stefan Meyer Institut f\"{u}r Subatomare Physik (SMI), Vienna, Austria\\
$^{102}$ SUBATECH, IMT Atlantique, Nantes Universit\'{e}, CNRS-IN2P3, Nantes, France\\
$^{103}$ Suranaree University of Technology, Nakhon Ratchasima, Thailand\\
$^{104}$ Technical University of Ko\v{s}ice, Ko\v{s}ice, Slovak Republic\\
$^{105}$ The Henryk Niewodniczanski Institute of Nuclear Physics, Polish Academy of Sciences, Cracow, Poland\\
$^{106}$ The University of Texas at Austin, Austin, Texas, United States\\
$^{107}$ Universidad Aut\'{o}noma de Sinaloa, Culiac\'{a}n, Mexico\\
$^{108}$ Universidade de S\~{a}o Paulo (USP), S\~{a}o Paulo, Brazil\\
$^{109}$ Universidade Estadual de Campinas (UNICAMP), Campinas, Brazil\\
$^{110}$ Universidade Federal do ABC, Santo Andre, Brazil\\
$^{111}$ University of Cape Town, Cape Town, South Africa\\
$^{112}$ University of Houston, Houston, Texas, United States\\
$^{113}$ University of Jyv\"{a}skyl\"{a}, Jyv\"{a}skyl\"{a}, Finland\\
$^{114}$ University of Kansas, Lawrence, Kansas, United States\\
$^{115}$ University of Liverpool, Liverpool, United Kingdom\\
$^{116}$ University of Science and Technology of China, Hefei, China\\
$^{117}$ University of South-Eastern Norway, Kongsberg, Norway\\
$^{118}$ University of Tennessee, Knoxville, Tennessee, United States\\
$^{119}$ University of the Witwatersrand, Johannesburg, South Africa\\
$^{120}$ University of Tokyo, Tokyo, Japan\\
$^{121}$ University of Tsukuba, Tsukuba, Japan\\
$^{122}$ University Politehnica of Bucharest, Bucharest, Romania\\
$^{123}$ Universit\'{e} Clermont Auvergne, CNRS/IN2P3, LPC, Clermont-Ferrand, France\\
$^{124}$ Universit\'{e} de Lyon, CNRS/IN2P3, Institut de Physique des 2 Infinis de Lyon, Lyon, France\\
$^{125}$ Universit\'{e} de Strasbourg, CNRS, IPHC UMR 7178, F-67000 Strasbourg, France, Strasbourg, France\\
$^{126}$ Universit\'{e} Paris-Saclay, Centre d'Etudes de Saclay (CEA), IRFU, D\'{e}partment de Physique Nucl\'{e}aire (DPhN), Saclay, France\\
$^{127}$ Universit\'{e}  Paris-Saclay, CNRS/IN2P3, IJCLab, Orsay, France\\
$^{128}$ Universit\`{a} degli Studi di Foggia, Foggia, Italy\\
$^{129}$ Universit\`{a} del Piemonte Orientale, Vercelli, Italy\\
$^{130}$ Universit\`{a} di Brescia, Brescia, Italy\\
$^{131}$ Variable Energy Cyclotron Centre, Homi Bhabha National Institute, Kolkata, India\\
$^{132}$ Warsaw University of Technology, Warsaw, Poland\\
$^{133}$ Wayne State University, Detroit, Michigan, United States\\
$^{134}$ Westf\"{a}lische Wilhelms-Universit\"{a}t M\"{u}nster, Institut f\"{u}r Kernphysik, M\"{u}nster, Germany\\
$^{135}$ Wigner Research Centre for Physics, Budapest, Hungary\\
$^{136}$ Yale University, New Haven, Connecticut, United States\\
$^{137}$ Yonsei University, Seoul, Republic of Korea\\
$^{138}$  Zentrum  f\"{u}r Technologie und Transfer (ZTT), Worms, Germany\\
$^{139}$ Affiliated with an institute covered by a cooperation agreement with CERN\\
$^{140}$ Affiliated with an international laboratory covered by a cooperation agreement with CERN.\\

\end{flushleft} 

\end{document}